\documentclass[a4paper, 11pt]{article}

\usepackage{xcolor}
\usepackage{setspace}

\usepackage{hyperref}
\definecolor{mycolor}{RGB}{0,88,204}
\hypersetup{
  colorlinks=true,
  linkcolor=mycolor,
  urlcolor=mycolor,
  citecolor=mycolor
}
\usepackage{amsmath}
\usepackage{geometry}

\usepackage{fancyhdr}

\fancypagestyle{plain}{%
  \fancyhf{}  
  \fancyhead[R]{\thepage}  
    
}

\usepackage{mathptmx}

\usepackage{titlesec}

\titleformat{\section}
  {\normalfont\large\centering}{\thesection.}{1em}{\MakeUppercase} 
\titleformat{\subsection}
  {\normalfont\centering}{\thesubsection.}{1em}{\MakeUppercase} 
  
  \titlespacing{\paragraph}{10pt}{0pt}{6pt}[0pt]

\usepackage{listings}
\usepackage[T1]{fontenc}
\usepackage[utf8]{inputenc}
\usepackage{amssymb}

\definecolor{keywordcolor}{rgb}{0.7, 0.1, 0.1}   
\definecolor{tacticcolor}{rgb}{0.0, 0.1, 0.6}    
\definecolor{commentcolor}{rgb}{0.4, 0.4, 0.4}   
\definecolor{symbolcolor}{rgb}{0.0, 0.1, 0.6}    
\definecolor{sortcolor}{rgb}{0.1, 0.5, 0.1}      
\definecolor{attributecolor}{rgb}{0.7, 0.1, 0.1} 

\lstnewenvironment{code}[1][]%
{
   \noindent\newline
   \minipage{\linewidth} 
   \vspace{0.5\baselineskip}
   \lstset{
 	frame = lrtb, 
 	rulecolor=\color{mycolor},
 	escapeinside={/*!}{!*/},
	language=lean, 
	aboveskip = 5mm,
	belowskip = 5mm,
	xleftmargin=2mm,
	xrightmargin=2mm,
	}
	}
{\endminipage\newline}
\lstnewenvironment{codeLong}[1][]%
{
   \lstset{
 	frame = lrtb, 
 	rulecolor=\color{mycolor},
 	escapeinside={/*!}{!*/},
	language=lean, 
	aboveskip = 5mm,
	belowskip = 5mm,
	xleftmargin=2mm,
	xrightmargin=2mm,
	}
	}
{}
\newcommand{\codeLink}[1]{
  \vspace{-0.5cm}\hfill\href{https://github.com/HEPLean/HepLean/blob/1b951994ae3d882242b02d23957ef1ee7fa05f3d/HepLean/#1}{(source)}
  }
 \newcommand{\textLink}[1]{\href{https://github.com/HEPLean/HepLean/blob/1b951994ae3d882242b02d23957ef1ee7fa05f3d/HepLean/#1}{source}}
 \newcommand{\textLinkB}[1]{\href{https://github.com/HEPLean/HepLean/blob/1b951994ae3d882242b02d23957ef1ee7fa05f3d/HepLean/#1}{(source)}}
\title{HepLean: Digitalising high energy physics}
\author{Joseph Tooby-Smith \\ \textit{Department of Physics, LEPP, Cornell University, Ithaca, NY 14853, USA}}
\date{\today}  

\begin{document}

\maketitle 
\vspace{-1cm}
\begin{abstract}
We introduce HepLean, an open-source project to digitalise definitions, theorems, proofs, and calculations in high energy physics using the interactive theorem prover Lean 4.  HepLean has the potential to benefit the high energy physics community in four ways: making it easier to find existing results, allowing the creation of new results using artificial intelligence and automated methods, allowing easy review of papers for mathematical correctness, and providing new ways to teach high energy physics. We will discuss these in detail.
We will also demonstrate the digitalisation of three areas of high energy physics in HepLean: Cabibbo-Kobayashi-Maskawa matrices in flavour physics, local anomaly cancellation, and Higgs physics. 
\end{abstract}

\begin{center}
Link to GitHub repository: \url{https://github.com/HEPLean/HepLean}.
\end{center}

\tableofcontents

\section{Introduction}

The aim of this paper is to introduce the reader to the GitHub hosted project  HepLean. 
The purpose of HepLean is to digitalise results (meaning definitions, theorems, proofs, and calculations) from high energy physics in a way that computers can read them systematically, and check their correctness. 

In HepLean, we carry out this digitalisation using the computer programming
language Lean~4 (simply `Lean' from hereon)~\cite{lean}. Lean is  a special type of computer language called an
interactive theorem prover. It allows you to write down definitions and theorems using its
mathematical foundation called `dependent type theory'. Lean also allows you to give proofs of
theorems using dependent type theory, and will certify if a proof is correct or not.
  
Results in high energy physics aren't typically, if ever, written using
dependent type theory (at least explicitly). So at this point the reader may be wondering if it is
possible to digitalise \emph{any} result from high energy physics into Lean. The reason it is
possible, is that the mathematicians and computer scientists have done most of the ground-work for
us in the project MathLib~\cite{mathlib}. 

MathLib is an ongoing project written in Lean to digitalise results in 
mathematics. It includes, for example, results from representation theory, the theory of smooth
manifolds, topology, category theory etc., all built up from the foundations of dependent type
theory. In Lean, you can use any theorem already proved, or any definition already made, in the proofs
of new theorems and the making of new definitions. Thus, to write results from high energy physics
in Lean, we can use the corpus of results in MathLib to help. One does not need to work with dependent type theory explicitly. 
Therefore, with MathLib at hand, the task of making definitions, proving theorems, or performing calculations (which can be reframed as theorems) in Lean, becomes much closer to what high energy physicists are familiar with. It is this work of the mathematicians and computer scientists which mean it is possible to
digitalise results  from high energy physics into Lean.

Using examples, we now give the reader a brief taste of what Lean code looks like. The two key objects in Lean are definitions, and theorems (or equivalently lemmas). A definition
has the following basic structure:
\begin{code}
def name_of_object (p1 : parameter1) ... : type_of_object := the_def_of_the_object
\end{code}
As an example, the definition of the fixed-point subset of a group action of \lstinline|G| on a set \lstinline|α| is written in MathLib (with some minor changes) as follows:
\begin{code}
def fixedBy (α : Type) (g : G) : Set α :=
  { x | g • x = x }
\end{code}

A lemma or theorem has the basic structure:
\begin{code}
 theorem name_of_theorem (p1 : parameter1) ... (a1 : assumption1) ... : 
 		thing_to_be_proved := by
	proof
\end{code}
For example, the theorem `under a group action, points fixed by the action of $g$  are also
 fixed by the action of $g^{-1}$' appears in MathLib as the following Lean code:
\begin{code}
theorem fixedBy_inv (g : G) : fixedBy α g⁻¹ = fixedBy α g := by
  ext
  rw [mem_fixedBy, mem_fixedBy] -- curser at end of this line (see below)
  rw [inv_smul_eq_iff, eq_comm]
\end{code}
Throughout this paper we will see more examples of definitions and theorems as well as some minor variations thereon.

When talking about results in  high energy physics  we also refer to calculations, for which there is no specific structure in Lean. We get around this by storing the \emph{result} of a calculation in Lean as the statement of a theorem, and the calculation itself as the proof of that theorem.

As you
work through a proof of a theorem in Lean, you are presented  with  information summarising the current goal to be proved, and the assumptions and parameters one has available. Lean presents this information in a `tactic state' which has the following basic structure:
\begin{code}
p1 : parameter1 
...
a1 : assumption1 
⊢ current_state_of_goal
\end{code}
As an example, the last-but-one code snippet gives a proof of a theorem `\lstinline|fixedBy_inv|'. For this theorem, when the computer curser is placed (in an appropriate code editor) where indicated, the tactic state shown is:
\begin{code}
α : Type u_1
G : Type u_2
inst✝³ : Group G
inst✝² : MulAction G α
g : G
x✝ : α
⊢ g⁻¹ • x✝ = x✝ ↔ g • x✝ = x✝
\end{code}
 The tactic state here tells us, for example, that we have a type $G$ which has an
instance as a group, and that we have a multiplicative action of $G$ on the type $\alpha$. 

Using previously proven theorems or automated proving tactics, one can change the current goal in the tactic state, or for example, add new parameters. To prove a theorem in Lean the current goal must be turned to True, or for example, a contradiction given among the inputted assumptions.

HepLean is not the first time Lean has been applied outside the
 ivory towers of the mathematicians. Notably, the paper~\cite{josephson} used Lean to digitalise
 results in absorption theory, thermodynamics, and kinematics. The authors of that paper proposed the
 idea of a library of results in Lean for all the sciences. HepLean is a step in that direction for
 high energy physics. Additionally, the package SciLean~\cite{SciLean}, aims to provide a framework for scientific computing in Lean, and allows, for example, simulations of the harmonic oscillator in Lean. There are also individuals,  notably Winston Yin, working on digitalising into MathLib areas of mathematics of specific interest to physicists.
 Interactive theorem provers other than Lean, have also been used to digitalise
 specific results of science,  for example~\cite{lu2017formalization} used the
 interactive theorem prover Coq to digitalise parts of  special relativity.
 
\subsection{Motivation}\label{sec:Motivation}
It is perhaps unsurprising that the motivation for HepLean is very similar to that of MathLib. The latter has been well-documented (see e.g., the Quanta article~\cite{Quanta}).  HepLean is motivated by its potential to benefit the high energy physics community. Here we will discuss four ways it does this:

\paragraph{Finding results in the literature:}

As a field, we currently store results from high energy physics in academic papers. This is a non-linear storage method in that results related to the same subject are scattered across the literature. This is unlike, for example, the storage of information in a text-book which is linear. Non-linear storage makes looking-up information hard. No doubt some readers of this paper will have struggled in the past to find a calculation or theorem in the high energy physics literature. HepLean, following MathLib, stores information linearly, meaning all results relating to e.g.,\ CKM matrices, are in the same place within the project. This, along with Lean-specific look-up tools such as `loogle'~\cite{loogle}, or the Lean command `\lstinline[language = {}]|exact?|', mean that HepLean would  make it easier to find results across high energy physics.

\paragraph{Creating new results:}
 HepLean can streamline the process of proving new
theorems or doing calculations.
 On top of the ability to easily use results previously digitalised in HepLean, the mathematicians and computer scientists have
developed some powerful automated proof tactics to help. These automated tactics are commands in Lean that automatically do
part of a calculation or proof. 
 Some of these are algorithmic. Two examples are  \lstinline[language = {}]|simp|  which is similar to \lstinline[language = {}]|Simplify| in Mathematica,
 and \lstinline[language = {}]|aesop| which automatically searches for simple proofs. Others depend on artificial intelligence to fill in parts of or complete proofs, see e.g.,~\cite{LeanDojo}.  The machine learning models these AIs are based on could be improved using
the corpus of results from HepLean as training data.
  
\paragraph{Reviewing for mathematical correctness:}
  
HepLean can make it easier to review papers for
mathematical correctness. As previously mentioned,  Lean will check if a proof is correct or not (it
will not `compile' if it contains an invalid proof of a result). Thus, for results written in HepLean
there is no question about whether they are mathematically correct or not. There cannot be, for
example, a mathematical typo, or a factor of 2 missing or a minus sign error. This is unlike LaTeX,
which will compile even with such errors.\footnote{ Proof: $-1+1=2$.} Putting results into HepLean
thus eliminates the need for a human to review the paper for mathematical correctness and helps
authors to be confident in the mathematical correctness of their own papers.

\paragraph{Pedagogy:}
 HepLean will provide at least two new
ways to teach high energy physics.  The first way is through Lean games. These games work by getting
the player to fill out a proof to a given theorem (which could be a calculation) and automatically
checking if the given proof is correct or not. These games can be created using the Lean 4 Game
code~\cite{LGS}. As an example in mathematics there is the  `Natural Numbers Game'~\cite{NNG} and
the `Set Theory Game`~\cite{STG} among others. The second way is through the creation of 
Master projects digitalising areas of high energy physics into HepLean. Students can make their own contributions to the field whilst learning about an area of high energy physics, and getting
instantaneous feedback through Lean on the correctness of their work. This approach has been used
extensively in MathLib.  

At the time of writing, HepLean has three areas of
high energy physics in which some results have been digitalised. These are
Cabibbo-Kobayashi-Maskawa (CKM) matrices in flavour physics, local anomaly cancellation, and Higgs
physics. In the next three sections we will present these digitalisations. The aim will be the broad
picture rather than the detail. These areas cover a range of mathematics common in high energy physics.
In particular in \S\ref{sec:CKMMatrix} we will see complex analysis and equivalence relations. In
\S\ref{sec:ACCs} we will see linear algebra, and group and representation theory. Lastly, in
\S\ref{sec:Higgs} we will see differential geometry (smooth maps) and extrema of functions. This
paper will conclude in \S\ref{sec:future} with a discussion of future work. 

\section{CKM Matrices}\label{sec:CKMMatrix}
A CKM matrix describes how quarks interact with the weak-force, and in particular govern flavour-changing weak interactions in particle physics. 

In HepLean, several results relating to CKM matrices have been digitalised by the author. Primarily  these are: the definition of a CKM matrix; relations between the elements of a CKM matrix; the equivalence relation on CKM matrices; invariants of CKM matrices; and, the standard parameterisation of a CKM matrix. We will go through these in turn, while simultaneously introducing various aspects of coding in Lean. 
\subsection{Definition} 

 A CKM matrix is a $3\times 3$  unitary matrix. To digitalise this information in HepLean we define the `type' (which can be thought of  as a `set') of CKM matrices as follows:\footnote{For each code snippet in this paper from HepLean we provide a link `(source)' to that code in a stable version of HepLean. For presentational purposes, small variations may exist between the code snippet appearing in this paper and that found following the link.}  
\begin{code}
/*!\codeLink{FlavorPhysics/CKMMatrix/Basic.lean\#L117}!*/
/-- The type of CKM matrices. -/
def CKMMatrix : Type := unitaryGroup (Fin 3) ℂ
\end{code}
Here  \lstinline|unitaryGroup| and the complex numbers \lstinline|ℂ| are imported from MathLib.
Types have members, roughly corresponding to elements of a set. In Lean this is written as
\lstinline|(V : CKMMatrix)|, which says that \lstinline|V| is a member of the type
\lstinline|CKMMatrix|. Thus, here it tells us (and Lean) that \lstinline|V| is a $3\times
3$-unitary matrix.

In physics, it is common to denote the elements of a CKM matrix $V$ as follows:
\begin{equation*}
	\left(\begin{smallmatrix}
 		V_{ud} & V_{us} & V_{ub} \\
		V_{cd} & V_{cs} & V_{cb}\\ 
		V_{td} & V_{ts} & V_{tb}\end{smallmatrix}\right).
\end{equation*}
In Lean one can define notation to closely match this. For instance, if  \lstinline|V| is a CKM matrix we can define the Lean notation \lstinline|[V]ud| as corresponding to the top left element of  \lstinline|V|. This is done as follows:
\begin{code}
/*!\codeLink{FlavorPhysics/CKMMatrix/Basic.lean\#L126}!*/
/-- The `ud`th element of the CKM matrix. -/
scoped[CKMMatrix] notation (name := ud_element) "[" V "]ud" => V.1 0 0
\end{code}
We can then use the notation \lstinline|[V]ud| in statements of definitions and theorems. We will see this used in the following subsections.

\subsection{Relations between elements} 

A CKM matrix, being unitary, has many relations between its elements. 
For example, each row is normalised to 1. 
HepLean contains these relations as short Lemmas. We do this so that it is easy to use these relationships in the proofs or calculations of more substantial results. 

To give an example, the lemma that the first row of a CKM matrix is normalised to~$1$ appears in HepLean as follows:
\begin{code}
/*!\codeLink{FlavorPhysics/CKMMatrix/Relations.lean\#L40}!*/
lemma fst_row_normalized_abs (V : CKMMatrix) : 
    abs [V]ud ^ 2 + abs [V]us ^ 2 + abs [V]ub ^ 2 = 1 := by
  ...
\end{code}
We omit the proof here, replacing it with `\ldots' (it can be found by following the `(source)' link given in the code snippet). In this theorem the notation `\lstinline|abs [V]ud|' indicates the absolute value of the complex number  \lstinline|[V]ud|. This notation, along with many properties of the complex numbers, is imported from MathLib.

It is worth repeating at this point that MathLib allows us to write results at a high-level, 
despite the dependent type theory foundations of Lean.

\subsection{Equivalence relation} 
We now turn to the equivalence relation defined on the type (or set) of CKM matrices. Two CKM matrices are equivalent if they are equal up-to phase-shifts in the quarks. In HepLean the underlying relation appears as:
\begin{code}
/*!\codeLink{FlavorPhysics/CKMMatrix/Basic.lean\#L70}!*/
/-- The equivalence relation between CKM matrices. -/
def phaseShiftRelation (U V : unitaryGroup (Fin 3) ℂ) : Prop :=
  ∃ a b c e f g , U = phaseShift a b c * V * phaseShift e f g
\end{code}
Here `\lstinline|phaseShift a b c|' corresponds to the unitary matrix $\mathrm{diag}(e^{ia}, e^{ib}, e^{ic})$~\textLinkB{FlavorPhysics/CKMMatrix/Basic.lean\#L70}.
In HepLean this is followed by a lemma, \lstinline|phaseShiftRelation_equiv|~\textLinkB{FlavorPhysics/CKMMatrix/Basic.lean\#L57},  stating that  \lstinline|phaseShiftRelation| is an equivalence relation.

With this equivalence relation we can tell Lean to treat \lstinline|CKMMatrix| as a setoid which is simply a type (or set) with a specified equivalence relation. This is done as follows:
\begin{code}
/*!\codeLink{FlavorPhysics/CKMMatrix/Basic.lean\#L152}!*/
instance CKMMatrixSetoid : Setoid CKMMatrix := 
    ⟨phaseShiftRelation, phaseShiftEquivRelation⟩
\end{code}
At this point  \lstinline|CKMMatrix| is now the structure of a Type with an equivalence relation. We can use, for example, the notation \lstinline|V ≈ U|  as a statement in Lean that \lstinline|V| and \lstinline|U| are to be treated as equivalent under this relation. Similarly, we can use `\lstinline|Quotient CKMMatrixSetoid|' for the type of equivalence classes of CKM matrices under the equivalence relation.

\subsection{Invariants}

An invariant of a CKM matrix is a function from the set of CKM matrices to, for example,\ the complex or real numbers which is well-defined on equivalence classes. There are a number of invariants of CKM matrices currently defined in HepLean. Here we will focus on one, the 
complex Jarlskog invariant~\cite{Jarlskog}. 

The complex Jarlskog invariant of a CKM matrix $V$ is $V_{us} V_{cb} V_{ub}^\ast V_{cs}^\ast$. In HepLean this is defined as follows: 
\begin{code}
/*!\codeLink{FlavorPhysics/CKMMatrix/Invariants.lean\#L31}!*/
/-- The complex jarlskog invariant for a CKM matrix. -/
def jarlskogℂCKM (V : CKMMatrix) : ℂ := [V]us * [V]cb * conj [V]ub * conj [V]cs
\end{code}
Here \lstinline|conj| is complex conjugation.  We note in Lean we could have equivalently defined it as a map \lstinline|CKMMatrix →  ℂ|, something we will see for \lstinline|jarlskogℂ| below. 

It is easy to check by hand that the complex Jarlskog invariant is well-defined on equivalence
 classes. In HepLean this result appears as the following lemma:
\begin{code}
/*!\codeLink{FlavorPhysics/CKMMatrix/Invariants.lean\#L36}!*/
lemma jarlskogℂCKM_equiv  (V U : CKMMatrix) (h : V ≈ U) :
    jarlskogℂCKM V = jarlskogℂCKM U := by
  obtain ⟨a, b, c, e, f, g, h⟩ := h
  change V = phaseShiftApply U a b c e f g  at h
  rw [h]
  simp only [jarlskogℂCKM, Fin.isValue, phaseShiftApply.ub,
  phaseShiftApply.us, phaseShiftApply.cb, phaseShiftApply.cs]
  simp [← exp_conj, conj_ofReal, exp_add, exp_neg]
  have ha : cexp (↑a * I) ≠ 0 := exp_ne_zero _
  have hb : cexp (↑b * I) ≠ 0 := exp_ne_zero _
  have hf : cexp (↑f * I) ≠ 0 := exp_ne_zero _
  have hg : cexp (↑g * I) ≠ 0 := exp_ne_zero _
  field_simp
  ring
\end{code}
This lemma takes as an input  two CKM matrices \lstinline|V| and  \lstinline|U| as well as the assumption (or proof)   \lstinline|h| that \lstinline|V ≈ U|. Note the proof of this lemma relies on the powerful tactics \lstinline|simp|, \lstinline|field_simp| and  \lstinline|ring| which automate much of the proof for us. 

We can encode the fact that the Jarlskog is well-defined on equivalence classes by redefining it as a map from `\lstinline|Quotient CKMMatrixSetoid|'  to the complex numbers `\lstinline|ℂ|'. In HepLean this is done as follows:
\begin{code}
/*!\codeLink{FlavorPhysics/CKMMatrix/Invariants.lean\#L53}!*/
/-- The complex jarlskog invariant for an equivalence class 
of CKM matrices. -/
def jarlskogℂ : Quotient CKMMatrixSetoid → ℂ :=
	Quotient.lift jarlskogℂCKM jarlskogℂCKM_equiv
\end{code}

\subsection{Standard parameterisation} 
Given four real numbers $\theta_{12}$, $\theta_{13}$, $\theta_{23}$ and $\delta_{13}$ one can write down the matrix
\begin{equation*}
	\left(\begin{smallmatrix}
    c_{12}c_{13} & s_{12}c_{13} & s_{13}e^{-i\delta_{13}} \\
    -s_{12}c_{23}-c_{12}s_{13}s_{23}e^{i\delta_{13}} & c_{12}c_{23}-s_{12}s_{13}s_{23}e^{i\delta_{13}} & s_{23}c_{13} \\
    s_{12}s_{23}-c_{12}s_{13}c_{23}e^{i\delta_{13}} & -c_{12}s_{23}-s_{12}s_{13}c_{23}e^{i\delta_{13}} & c_{23}c_{13}
    \end{smallmatrix}\right)
\end{equation*}
where $c_{12} = \cos \theta_{12}$, $S_{12} = \sin \theta_{12}$ etc. This is a $3\times 3$ unitary matrix, and thus represents a CKM matrix. A CKM matrix written in this form is in the `standard parameterisation'~\cite{CKMStandardParam}. 

In HepLean this matrix (as a matrix and not a CKM matrix) is defined as follows:
\begin{code}
/*!\codeLink{FlavorPhysics/CKMMatrix/StandardParameterization/Basic.lean\#L28}!*/
/-- Given four reals `θ₁₂ θ₁₃ θ₂₃ δ₁₃` the standard 
paramaterization of the CKM matrixas a `3×3` complex matrix. -/
def standParamAsMatrix (θ₁₂ θ₁₃ θ₂₃ δ₁₃ : ℝ) : Matrix (Fin 3) (Fin 3) ℂ  :=
  ![![Real.cos θ₁₂ * Real.cos θ₁₃, Real.sin θ₁₂ * Real.cos θ₁₃, 
   Real.sin θ₁₃ * exp (-I * δ₁₃)],
    ![(-Real.sin θ₁₂ * Real.cos θ₂₃) - (Real.cos θ₁₂ * Real.sin θ₁₃ * Real.sin θ₂₃ 
      * exp (I * δ₁₃)), Real.cos θ₁₂ * Real.cos θ₂₃ - Real.sin θ₁₂ * Real.sin θ₁₃ * 
      Real.sin θ₂₃ * exp (I * δ₁₃), Real.sin θ₂₃ * Real.cos θ₁₃],
    ![Real.sin θ₁₂ * Real.sin θ₂₃ - Real.cos θ₁₂ * Real.sin θ₁₃ * Real.cos θ₂₃ 
      * exp (I * δ₁₃), (-Real.cos θ₁₂ * Real.sin θ₂₃) - (Real.sin θ₁₂ * Real.sin θ₁₃ 
      * Real.cos θ₂₃ * exp (I * δ₁₃)), Real.cos θ₂₃ * Real.cos θ₁₃]]
\end{code}
To lift it to a CKM matrix we must show it is unitary. In HepLean, we do this in a lemma called \lstinline|standParamAsMatrix_unitary|~\textLinkB{FlavorPhysics/CKMMatrix/StandardParameterization/Basic.lean\#L41}.
We can then define the standard parameterisation as a CKM matrix as follows:
\begin{code}
/*!\codeLink{FlavorPhysics/CKMMatrix/StandardParameterization/Basic.lean\#L101}!*/
/-- Given four reals `θ₁₂ θ₁₃ θ₂₃ δ₁₃` the standard paramaterization 
of the CKM matrix as a CKM matrix. -/
def standParam (θ₁₂ θ₁₃ θ₂₃ δ₁₃ : ℝ) : CKMMatrix :=
  ⟨standParamAsMatrix θ₁₂ θ₁₃ θ₂₃ δ₁₃, by
   rw [mem_unitaryGroup_iff']
   exact standParamAsMatrix_unitary θ₁₂ θ₁₃ θ₂₃ δ₁₃⟩
\end{code}

The standard parameterisation is useful because it is true that every CKM matrix up to equivalence can be written in the standard parameterisation. In HepLean this appears as the following theorem:
\begin{code}
/*!\codeLink{FlavorPhysics/CKMMatrix/StandardParameterization/StandardParameters.lean\#L674}!*/
theorem exists_for_CKMatrix (V : CKMMatrix) :
    ∃ (θ₁₂ θ₁₃ θ₂₃ δ₁₃ : ℝ), V ≈ standParam θ₁₂ θ₁₃ θ₂₃ δ₁₃ := by
  ...
\end{code}
We have omitted the proof since it depends on a number of lemmas not given here.

In this section we have presented how results from CKM matrices are digitalised in HepLean. We have seen the use of complex numbers and equivalence relations in Lean, and gone through some basic aspects of coding in Lean.

\section{Anomaly cancellation}\label{sec:ACCs}

When extending a gauge theory by a $\mathfrak{u}(1)$ Lie algebra factor, each fermion gains a charge with respect to that  $\mathfrak{u}(1)$. If the $\mathfrak{u}(1)$  comes from a $U(1)$ Lie group, then these charges can be scaled to rational numbers. For a consistent theory the charges cannot be chosen arbitrarily. In particular, they must satisfy a series of polynomial equations arising from triangle Feynman diagrams, which ensure the cancellation of local anomalies. We call these equations the anomaly cancellation conditions (ACCs). 

HepLean has the digitalisation of the main results of a series of papers in the area of local anomaly cancellation. Firstly, HepLean contains the parameterisation of the solutions to the ACCs for a pure $U(1)$-gauge theory with an even (see e.g.,~\textLink{AnomalyCancellation/PureU1/Even/Parameterization.lean\#L121}) and odd number of fermions (see e.g,~\textLink{AnomalyCancellation/PureU1/Odd/Parameterization.lean\#L120})~\cite{CDF19, Allanach:2019gwp}, for a $U(1)$-extension to the Standard Model (SM) with three right handed neutrinos (see e.g.,~\textLink{AnomalyCancellation/SMNu/PlusU1/QuadSolToSol.lean\#L143})~\cite{Allanach:2020zna},  and for a $U(1)$-extension to the Minimal Supersymmetric Standard Model with three right handed neutrinos (see e.g.,~\textLink{AnomalyCancellation/MSSMNu/OrthogY3B3/ToSols.lean\#L450})~\cite{Allanach:2021yjy}. Secondly, it contains the main result of~\cite{LT19} which will be discussed in~\S\ref{sec:LohitsiriTong}. 

In this section, for illustration purposes, we will focus our attention on a small generalisation of the set-up of~\cite{LT19}. Specifically,  the
extension of the  $\mathfrak{su}(3)\times \mathfrak{su}(2)$ gauge algebra of the SM to $\mathfrak{su}(3)\times \mathfrak{su}(2) \times \mathfrak{u}(1)$ assuming $n$-families of SM fermions.
 We will summarise some definitions involved, discuss permutations among families, and finish by briefly discussing the main result of~\cite{LT19} which holds for the $n=1$ case. In this section we will see the use of group actions, representation theory, and linear maps in Lean. 

\subsection{Definitions}
To extend $\mathfrak{su}(3)\times \mathfrak{su}(2)$  to  $\mathfrak{su}(3)\times \mathfrak{su}(2) \times \mathfrak{u}(1)$ we need to associate to each fermion of the SM a rational charge.  
The set of all possible rational charges forms a $(5\times n)$-dimensional vector-space over the rationals. 
The $5$ comes from the five species of the SM $Q$, $U$, $D$, $E$, and $L$, and $n$ comes from the $n$-families (so e.g., we have $n$ copies of $Q$, one in each family). In HepLean this vector space appears as \lstinline|(SMCharges n).charges| 
(see e.g.,~\textLink{AnomalyCancellation/SM/Basic.lean\#L24} and~\textLink{AnomalyCancellation/Basic.lean\#L36}). Given an element  \lstinline|(S : (SMCharges n).charges)| we define the notation `\lstinline|Q S i|'~\textLinkB{AnomalyCancellation/SM/Basic.lean\#L65} for the charge associated to the left-handed quark in the $i$th family, and similar for other species involved. 

One of the anomaly equations the charges must satisfy is the $ \mathfrak{su}(2)$-anomaly equation. This is a polynomial equation of degree 1 in the charges. In HepLean it is defined as a linear map (denoted ` \lstinline|→l[ℚ]|') from the vector space \lstinline|(SMCharges n).charges| to \lstinline|ℚ| as follows:
\begin{code}
/*!\codeLink{AnomalyCancellation/SM/Basic.lean\#L118}!*/
/-- The `su(2)` anomaly equation. -/
def accSU2 : (SMCharges n).charges →l[ℚ] ℚ where
  toFun S := ∑ i, (3 * Q S i + L S i)
  map_add' S T := by
    simp only
    repeat rw [map_add]
    simp [Pi.add_apply, mul_add]
    repeat erw [Finset.sum_add_distrib]
    ring
  map_smul' a S := by
    simp only
    repeat erw [map_smul]
    simp [HSMul.hSMul, SMul.smul]
    repeat erw [Finset.sum_add_distrib]
    repeat erw [← Finset.mul_sum]
    ring
\end{code}
Note, to define \lstinline|accSU2| as a linear map, we have to provide the function of sets \lstinline|toFun| and prove that it is linear with respect to addition \lstinline|map_add'| and scalar multiplication \lstinline|map_smul'|. A set of charges respects the $\mathfrak{su}(2)$-ACC if it is in the kernel of the map \lstinline|accSU2|.

The $\mathfrak{su}(2)$ anomaly equation is not the only ACC which must be satisfied by the charges.
There are also the linear gravitational ACC defined in HepLean as the linear map \lstinline|accGrav|~\textLinkB{AnomalyCancellation/SM/Basic.lean\#L88}; the
linear $\mathfrak{su}(3)$-ACC defined as the linear map  \lstinline|accSU3|~\textLinkB{AnomalyCancellation/SM/Basic.lean\#L148}; and, the non-linear
cubic ACC associated to a Feynman diagram with three  $\mathfrak{u}(1)$'s. HepLean denotes this latter ACC 
 as \lstinline|accCube| and defines it as a $\mathbb{Q}$-equivariant map from
\lstinline|(SMCharges n).charges| to \lstinline|ℚ|, where $a\in \mathbb{Q}$ acts on the former vector
space by scalar multiplication and on the latter by $b\mapsto a^3 b$~\textLinkB{AnomalyCancellation/SM/Basic.lean\#L306}.

\subsection{Permutations among species}

There is a permutation group acting on the vector space of charges given by the permutation between families (or equivalently among species). For our set-up this group is $S_n^{\times 5}$ where $S_n$ is the permutation group of $n$-objects. In HepLean, we define this group as follows:
\begin{code}
/*!\codeLink{AnomalyCancellation/SM/Permutations.lean\#L27}!*/
/-- The group of `Sₙ` permutations for each species. -/
def permGroup (n : ℕ) := ∀ (_ : Fin 5),  Equiv.Perm (Fin n)
\end{code}

This group acts on the vector-space of charges in the natural way. In HepLean this action is defined via a representation as follows:
\begin{code}
/*!\codeLink{AnomalyCancellation/SM/Permutations.lean\#L55}!*/
/-- The representation of `(permGroup n)` acting on the vector space 
of charges. -/
def repCharges {n : ℕ} : Representation ℚ (permGroup n) (SMCharges n).charges where
	toFun :=...
	map_mul' f g := by ...
	map_one' := by ...
\end{code}
This uses the type `\lstinline|Representation _ _ _|' defined in MathLib. To define this representation we have to provide a function from the group to the linear maps of the vector space, and show that this obeys the usual properties of representation with respect to multiplication \lstinline|map_mul'| and the identity \lstinline|map_one'|. 

All four ACCs are invariant under this group action. As an example, the lemma that the $\mathfrak{su}(2)$-ACC is invariant appears in HepLean as follows: 
\begin{code}
/*!\codeLink{AnomalyCancellation/SM/Permutations.lean\#L96}!*/
lemma accSU2_invariant (f : permGroup n) (S : (SMCharges n).charges)  :
    accSU2 (repCharges f S) = accSU2 S :=
  accSU2_ext
    (by simpa using toSpecies_sum_invariant 1 f S)
\end{code}

Again it is important to reiterate that once proved, these theorems can be used to help prove more complicated results. 

\subsection{Excluding the gravitational anomaly}\label{sec:LohitsiriTong}

In~\cite{LT19} the authors looked at the $n=1$ case of extending the $\mathfrak{su}(3)\times
\mathfrak{su}(2)$ gauge algebra of the SM to $\mathfrak{su}(3)\times \mathfrak{su}(2)
\times \mathfrak{u}(1)$ assuming $n$-families of SM fermions. This corresponds to 1-family, or
equivalently the family-universal scenario. They studied solutions to the ACCs excluding the
gravitational one. In HepLean we define  `\lstinline|SMNoGrav n|'~\textLinkB{AnomalyCancellation/SM/NoGrav/Basic.lean\#L21} which is a member of a type called
\lstinline|ACCSystem|~\textLinkB{AnomalyCancellation/Basic.lean\#L219}. Here, all we need to know is that  `\lstinline|(SMNoGrav n).Sols|'  is the type
of solutions to the $\mathfrak{su}(2)$, $\mathfrak{su}(3)$ and cubic ACCs.

The authors of~\cite{LT19}  show that any solution to these three ACCs automatically satisfies the fourth ACC, the gravitational one (assuming rationality of charges as done here). In HepLean this appears as follows: 
\begin{code}
/*!\codeLink{AnomalyCancellation/SM/NoGrav/One/Lemmas.lean\#L70}!*/
/-- Any solution to the ACCs without gravity satisfies
 the gravitational ACC.  -/
theorem accGravSatisfied {S : (SMNoGrav 1).Sols} (FLTThree : FermatLastTheoremWith ℚ 3) :
    accGrav S.val = 0 := by
  by_cases hQ : Q S.val (0 : Fin 1)= 0
  exact accGrav_Q_zero hQ
  exact accGrav_Q_neq_zero hQ FLTThree
\end{code}
The proof of this result is interesting since it ultimately depends on Fermat's last theorem for exponent $3$.  At the time of writing, a proof of this is not present in MathLib, and therefore is added as an assumption to the theorem.
\section{Higgs Physics}\label{sec:Higgs}
Currently, in HepLean there are a number of results related to the SM Higgs boson, and scalar potential. 
In this section we present a broad overview of this digitisation. 
We will start by defining the target vector space of the Higgs boson, followed by discussing the minimisation of the potential, defining generic Higgs fields, and finally discussing the smoothness properties of the potential as a function of spacetime.

\subsection{Target vector space of the Higgs}
The SM Higgs boson takes values in the vector space $\mathbb{C}^2$.
This vector space comes naturally equipped with the structure of an inner-product that will be useful in defining the Higgs potential. In HepLean we call this vector space \lstinline|higgsVec| and define it as follows:
\begin{code}
/*!\codeLink{StandardModel/HiggsField.lean\#L39}!*/
/-- The complex vector space in which the Higgs field takes values. -/
abbrev higgsVec := EuclideanSpace ℂ (Fin 2)
\end{code}
In MathLib there is an instance of an inner-product structure on  \lstinline|EuclideanSpace ℂ (Fin 2)|.
The use of  \lstinline|abbrev| in our definition above, rather than \lstinline|def|, 
induces for free an inner-product structure on \lstinline|higgsVec| from that on
 \lstinline|EuclideanSpace ℂ (Fin 2)|

\subsection{Minimising the potential}\label{sec:MinPotential}
The vector space \lstinline|higgsVec| has another possible interpretation. It is the vector space of those Higgs fields which are constant with respect to spacetime. For one of these fields $\phi$ the SM scalar potential is a number in $\mathbb{R}$ (rather than a map from spacetime to $\mathbb{R}$) given by:
\begin{equation*}
V = - \mu^2 |\phi|^2 + \lambda |\phi|^4,
\end{equation*}
for $\mu^2$ and $\lambda$ real parameters.
HepLean defines this scalar potential as follows:
 \begin{code}
 /*!\codeLink{StandardModel/HiggsField.lean\#L114}!*/
 /-- The higgs potential for `higgsVec`, i.e. for constant higgs fields. -/
def potential (μSq lambda : ℝ) (φ : higgsVec) : ℝ := - μSq  * ‖φ‖ ^ 2  +
  lambda * ‖φ‖ ^ 4	
 \end{code}

In the case when $\mu^2$ is negative, the minimum of this potential occurs at $\phi = 0$. This tells us that the vacuum-expectation value of the Higgs boson for these values of $\mu^2$ is zero, and no symmetry-breaking occurs. In HepLean, this property of the potential   appears as the following lemma:
\begin{code}
 /*!\codeLink{StandardModel/HiggsField.lean\#L265}!*/
lemma IsMinOn_potential_iff_of_μSq_nonpos {μSq lambda : ℝ} 
    (hLam : 0 < lambda) (hμSq : μSq ≤ 0) :
    IsMinOn (potential μSq lambda) Set.univ φ ↔ φ = 0 := by
   ...
\end{code}
Note that the proposition in this lemma  `\lstinline|IsMinOn (potential μSq lambda) Set.univ φ|' is defined only to be true when \lstinline|φ| is a global (due to \lstinline|Set.univ|) minimum of the potential.

In the case when $\mu^2$ is positive, the minimum of this potential occurs at non-zero $\phi$. This tells us that the vacuum-expectation value of the Higgs boson for this phase is non-zero, and that we have symmetry-breaking. In HepLean, this property of the potential appears as the following lemma:
\begin{code}
 /*!\codeLink{StandardModel/HiggsField.lean\#L250}!*/
lemma IsMinOn_potential_iff_of_μSq_nonneg {μSq lambda : ℝ} 
    (hLam : 0 < lambda) (hμSq : 0 ≤ μSq) :
    IsMinOn (potential μSq lambda) Set.univ φ ↔ ‖φ‖ ^ 2 = μSq /(2 * lambda) := by
   ...
\end{code}
 
\subsection{The Higgs field}
Spacetime for the SM is $\mathbb{R}^4$ as a manifold. In HepLean we define spacetime as:
\begin{code}
 /*!\codeLink{StandardModel/Basic.lean\#L25}!*/
/-- The space-time -/
abbrev spaceTime := EuclideanSpace ℝ (Fin 4)
\end{code}
Defining \lstinline|spaceTime| using MathLib's \lstinline|EuclideanSpace| allows us to easily put a smooth structure on it. A Minkowski metric could be added latter, but is not needed here. 

Higgs fields generically (i.e., not just constant ones) are most properly defined as smooth sections of the trivial vector bundle $\mathbb{R}^4 \times \mathbb{C}^2 \to \mathbb{R}^4$, which simply corresponds to smooth maps from spacetime to $\mathbb{C}^2$. This vector bundle in HepLean  is defined as follows:
\begin{code}
 /*!\codeLink{StandardModel/HiggsField.lean\#L42}!*/
/-- The trivial vector bundle 𝓡² × ℂ².-/
abbrev higgsBundle := Bundle.Trivial spaceTime higgsVec
\end{code}

HepLean defines the type of Higgs fields to be the type of smooth sections of this vector bundle as follows: 
\begin{code}
 /*!\codeLink{StandardModel/HiggsField.lean\#L48}!*/
/-- A higgs field is a smooth section of the higgs bundle. -/
abbrev higgsField : Type := SmoothSection (𝓡 4) higgsVec higgsBundle
\end{code}

\subsection{Smoothness of the potential}
In \S\ref{sec:MinPotential} we discussed the Higgs potential for constants Higgs fields and found its minima. Here we will discuss the Higgs potential for generic Higgs fields and its smoothness. 

For a Higgs field $\phi : \mathbb{R}^4 \to \mathbb{C}^4$ the potential at $x \in  \mathbb{R}^4$ is given by:
\begin{equation*}
  V (x) = - \mu^2 |\phi(x)|^2 + \lambda |\phi(x)|^4.
  \end{equation*}
This is a smooth map on spacetime.

In HepLean we first define the map $\mathbb{R}^4 \to \mathbb{R} : x \mapsto |\phi(x)|^2$ as follows:
\begin{code}
 /*!\codeLink{StandardModel/HiggsField.lean\#L402}!*/
/-- Given a `higgsField`, the map `spaceTime → ℝ` obtained 
by taking the square norm of the higgs vector.  -/
def normSq (φ : higgsField) : spaceTime → ℝ := fun x => ( ‖φ x‖ ^ 2)
\end{code}

Since \lstinline|φ| is smooth we expect this map to be smooth, but as it is defined here, Lean only knows it is a map from one type (set) to another. We must demonstrate to Lean that it is smooth with respect to the appropriate smooth structures on spacetime and on $\mathbb{R}$. This is done using the following lemma:
\begin{code}
 /*!\codeLink{StandardModel/HiggsField.lean\#L417}!*/	
lemma normSq_smooth (φ : higgsField) : 
    Smooth 𝓘(ℝ, spaceTime) 𝓘(ℝ, ℝ) φ.normSq := by
  rw [normSq_expand]
  refine Smooth.add ?_ ?_
  simp only [mul_re, conj_re, conj_im, neg_mul, sub_neg_eq_add]
  refine Smooth.add ?_ ?_
  refine Smooth.smul ?_ ?_
  exact φ.apply_re_smooth 0
  exact φ.apply_re_smooth 0
  refine Smooth.smul ?_ ?_
  exact φ.apply_im_smooth 0
  exact φ.apply_im_smooth 0
  simp only [mul_re, conj_re, conj_im, neg_mul, sub_neg_eq_add]
  refine Smooth.add ?_ ?_
  refine Smooth.smul ?_ ?_
  exact φ.apply_re_smooth 1
  exact φ.apply_re_smooth 1
  refine Smooth.smul ?_ ?_
  exact φ.apply_im_smooth 1
  exact φ.apply_im_smooth 1
\end{code}
As can be seen, the proof of this lemma relies on a series of properties of smoothness already encoded into MathLib. For example, \lstinline|Smooth.add| is the lemma that the addition of two smooth functions is also smooth. 

In HepLean we then define the Higgs potential for a generic Higgs field as follows:
\begin{code}
 /*!\codeLink{StandardModel/HiggsField.lean\#L443}!*/
/-- The Higgs potential of the form `- μ² * |φ| ² + λ * |φ| ⁴`. -/
def potential (φ : higgsField) (μSq lambda : ℝ ) (x : spaceTime) :  ℝ :=
  - μSq * φ.normSq x + lambda * φ.normSq x * φ.normSq x
\end{code}
Note that this now has a point in spacetime as a parameter, unlike for the potential for constant Higgs fields in \S\ref{sec:MinPotential}. 

 The potential for fixed \lstinline|φ|, \lstinline|μSq| and \lstinline|lambda| is (as far as Lean knows) only a map of types from spacetime to $\mathbb{R}$.  We thus provide a lemma in HepLean showing that it is also smooth as follows:
\begin{code}
/*!\codeLink{StandardModel/HiggsField.lean\#L448}!*/
lemma potential_smooth (φ : higgsField) (μSq lambda : ℝ) :
    Smooth 𝓘(ℝ, spaceTime) 𝓘(ℝ, ℝ) (fun x => φ.potential μSq lambda x) := by	
  simp only [potential, normSq, neg_mul]
  exact Smooth.add
    (Smooth.neg (Smooth.smul smooth_const φ.normSq_smooth))
    (Smooth.smul (Smooth.smul smooth_const φ.normSq_smooth) φ.normSq_smooth)
\end{code}

\section{Future work}\label{sec:future}
HepLean is an ongoing project. The long-term goal is to digitalise the whole of high energy physics within HepLean. With the development of new results in high energy physics, this long-term goal is potentially unreachable as the goalpost is ever moving.  Thus, in this section we will discuss the author's short-term goals for HepLean. 

The first goal is to digitalise some results from the two-Higgs doublet model (2HDM). In particular, we will digitalise aspects of the 2HDM scalar potential, similar to that of the Higgs field, discussed in \S\ref{sec:Higgs}. This project should be fairly straightforward, and will provide an explicit example of a digitalisation of beyond-the-standard-model physics. 

The second goal is to digitalise some aspects of fermions and more generic Lorentz group representations. In particular, the aim will be to set up notation and definitions in HepLean to make it easier to write down invariants of the Lorentz group in a style that high energy physicists are familiar with.  

The third goal is to digitalise some aspects of the theory of generalised symmetries. Thus far, the
material in HepLean leans towards the phenomenology side of the field. Digitalising aspects of
generalised symmetries will give an example of more formal areas in HepLean.

The fourth goal is to digitalise some aspects of experimental data. In HepLean one should be able to compare theoretical predictions with experimental data. This project will involve understanding the best way to write experimental data in Lean, and give an example in HepLean of a comparison of theory to experiment. A particular candidate for this is CKM matrices, which  has  results digitalised into HepLean as discussed in \S\ref{sec:CKMMatrix}. 

The fifth goal is to work on increasing the collaborative nature of HepLean, and getting more people involved. We plan to do this  by creating a high energy physics Lean game in order to teach physicists how to write Lean code. (On a side note, as a teaching resource, we plan to make Lean games for different areas of high energy physics.) Another way we plan to increase the collaborative nature of HepLean is by making a list of theorems and results to be digitalised, each graded by their expected difficulty. We hope this will make it easier for physicists to jump into digitalisation.

With that, if you are interested in contributing to HepLean, please feel free to either make a pull-request on the HepLean GitHub, or connect on the Lean Zulip channel (an online forum where most discussions about Lean and MathLib take place). 

\section*{Acknowledgments}

I thank  the  members of the Lean Zulip  for helpful discussions, with particular thanks to Tyler Josephson. I thank Margarita Gavrilova and Yuval Grossman for discussions about digitalising  aspects of flavour physics. I also thank  Csaba Cs\'aki,  Thomas Hartman, Baur Mukhametzhanov, and Maximilian Ruhdorfer for helpful discussions about digitalising of physics in general. The author is supported by the U.S. National Science Foundation (NSF) grant PHY-2014071.

\bibliographystyle{unsrturl}
\begin{spacing}{0.5}
\bibliography{MyBib}
\end{spacing}

\end{document}